\documentclass[preprint
amsmath,
amssymb,
aps,
pra,
floatfix,
]{revtex4-2}

\usepackage{graphicx}
\usepackage{dcolumn}
\usepackage{bm}
\usepackage{amsmath}

\begin{document}

%





\title{Multiorbital effects in high-order harmonic emission from CO$_2$}

\author{Lauren Bauerle, Yuqing Xia, Andres Mora and
Agnieszka Jaron}
\address{
 JILA and Department of Physics, University of Colorado, Boulder, CO 80309-0440, USA
}

\begin{abstract}
We study the ellipticity of high-order harmonics emitted from CO$_2$ molecule driven by linearly polarized laser fields using numerical simulations within the time-dependent density functional theory. We find that the overall ellipticity of the harmonics is small, which is in agreement with experimental data. On the other hand, our analysis of the numerical results indicates that several valence orbitals contribute significantly to the harmonic emission and some of these contributions show a strong ellipticity of the harmonics. The small ellipticity in the total harmonics signal arises from a combination of factors, namely,
the fact that harmonic emission from different orbitals is strongest at different alignment angles of the molecular axis with respect to the laser polarization direction, as well as interference effects and a strong laser coupling between several inner valence orbitals.
\end{abstract}


\maketitle

\section{\label{sec:intro}Introduction}

Ultrafast intense laser pulses can be a powerful tool for controlling and studying atomic and molecular structure and dynamics. For example manipulation of the electron dynamics in a molecule can be achieved by controlling the properties of the molecular wavepackets  \cite{review1}. In the case of probing performed by ultrashort laser pulses one of the advantages is that one does not have to consider the effect of the modifications of the potential energy surface (PES) topology by an electric field of the laser pulse. Ultrashort pulses allow for focusing studies on the ultrafast electron dynamics without adding the complexity of modification of the nuclear dynamics and consideration of laser-induced changes of the reaction even for high-intensity laser pulses. In particular, few femtosecond FWHM laser pulses allow for studies of ultrafast effects related directly to electron rearrangement during chemical reactions \cite{Li2010}, detailed studies of electron excitations \cite{Woerner2017}, and monitoring of ionization \cite{Li2010,Kupper2021,Garg2024}, all while the modification of the PES during laser-molecule interaction can be neglected due to the difference in the timescales between the dynamics of electrons and nuclei. 

Among intense laser field induced processes one prominent example is High-order harmonic generation (HHG), one of the highly nonlinear, nonperturbative processes \cite{McPherson1987,Ferray1988}. It results from the distortion of the electron density in the presence of the strong electromagnetic field of the laser and the power spectrum of the emitted harmonic radiation corresponds to the Fourier transform of the electron dipole acceleration. The intensity spectra of the emitted high harmonics shows some general characteristic features, such as a fast decrease of the signal over the first few harmonics followed by a region with fairly constant plateau harmonic intensities ending by a sharp cutoff, beyond which the harmonic intensity drops quickly.

Over the past few decades HHG has been an active area of research since it provides a source for coherent short-wavelength light, extending into the soft-X-ray regime \cite{Popmintchev2012}, and for ultrashort laser pulses and waveforms in the attosecond \cite{Hentschel2001,Paul2001}. 
Furthermore, it has been shown that HHG spectra contain information about the atomic and electronic structure of the target (e.g., \cite{Lein2002,Itatani2004,Woerner2009,Haessler2010,Vozzi2011,Schoun2014}), ultrafast molecular and intra-molecular electron dynamics (e.g., \cite{Baker2006,Li2008,Blaga2012,Miller2016}) as well as time resolution of chemical processes (e.g., \cite{Woerner2010}). 

Basic intuitive picture of HHG is provided by the semiclassical three-step model \cite{Schafer1993,Corkum1993,Lewenstein1994}, according to which an electron tunnels through the barrier created by the laser field and the Coulomb field into the continuum, is accelerated by the electric field of the laser first away from and then back to the parent ion. Upon return it recombines, emitting excess energy in the form of high-order harmonic radiation. Since the process occurs every half cycle of the driving laser field, an attosecond pulse train is produced. In recent years, it has been shown that, in particular for high-order harmonic generation from molecules, the generated harmonic spectra incorporate more features than predicted by the basic three-step single-active-electron model. One example are polarimetry measurements of high-order harmonic emission from aligned diatomic and linear triatomic molecules driven by linearly polarized laser fields. Surprisingly, strong elliptically polarized harmonics were observed for N$_2$ \cite{Zhao2009,Mairesse2010}, while in contrast CO$_2$ exhibited a much lower ellipticity in the harmonic emission \cite{Zhao2009}. Structural effects \cite{Etches2010,Le2010,Ramakrishna2010,vanderZwan2010,Son2010}, such as the symmetry of the Highest Occupied Molecular Orbital (HOMO) as well as interference effects, and ultrafast multielectron dynamics involving lower-lying orbitals in the molecule \cite{Xia2014} or in the molecular ion \cite{Mairesse2010} have been put forward as potential origins for the observed ellipticity.

In this article we focus on the role of multielectron and multiorbital effects in the neutral CO$_2$ molecule on the polarization state of high-order harmonics. We have shown previously \cite{Xia2014}, that results based on the time-dependent density functional theory (TDDFT) are in excellent agreement with the experimental data for N$_2$ \cite{Zhao2009,Mairesse2010}, if contributions from at least three Kohn-Sham orbitals are taken into account. Similar strong  influence of inner shell contributions has been observed and predicted for other strong-field processes as well \cite{Gibson1991,Talebpour1999,Becker2001,Jaron2004,McFarland2008,Le2009,Smirnova2009,Diveki2012,Xia2016,Smith2018}. 

Our results of numerical TDDFT simulations show that indeed the contributions from several valence orbitals contribute to the higher-order harmonic emission from CO$_2$. Moreover, we find that the emission from each of the orbitals is elliptically polarized. 
However, our results for the total high-order harmonic spectrum, which includes the contributions of up to six orbitals, surprisingly shows, in agreement with the experimental data \cite{Zhao2009}, almost no ellipticity. Thus, despite the fact that high-order harmonic generation from CO$_2$ appears to be a multielectron process with several orbitals actively involved, signatures in the ellipticity of the harmonic emission from the different orbitals fade away in the total signal. 

The article is organized as follows: In the next section we briefly outline the basics of the time-dependent density functional approach used for our numerical simulations. We then discuss the application to calculations of the ellipticity of high-order harmonic generation of molecules, including the proper account of the distribution of alignment in the molecular ensemble. Next, we compare the results of our calculations with the experimental data and analyze the contributions from the different valence orbitals to the total harmonic spectra. We end with a brief summary of our results. 

\section{Theory}

In the nonperturbative intensity regime the theoretical study of the interaction of multielectron targets, e.g.\ molecules, with ultrashort laser pulses is challenging. An approximative approach to analyze multielectron and multiorbital effects in strong-field processes utilizes the framework of the time-dependent density functional theory (TDDFT). In this section we outline the application of TDDFT to the calculation of high-harmonic generation in molecules, focusing in particular on the evaluation of the ellipticity of the radiation in an ensemble of aligned molecules. 

\subsection{TDDFT for strong-field induced molecular processes}

The TDDFT approach is based on the one-to-one correspondence between the time-dependent electron density $\rho({\bf r}, t)$ and the time-dependent potential in multielectron Schr\"odinger equation \cite{Runge1984}. The density is calculated from the time-dependent multielectron Schr\"odinger equation expressed as system of auxiliary time-dependent noninteracting single-electron Kohn-Sham equations: 
\begin{equation}
i \frac{\partial}{\partial t} \phi_k({\bf r}, t) 
= 
\left[ -\frac{\nabla^2}{2} + V_{KS}({\bf r}, t) \right] \phi_k({\bf r}, t)
\label{kohn}
\end{equation}
with 
\begin{equation}
\rho({\bf r}, t) = \sum\limits_{k=1}^{n} f_k |\phi_k({\bf r}, t)|^2
\end{equation}
where ${\bf r}$ is the electronic coordinate, $f_k$ is the electron population in the $k$-th Kohn-Sham orbital $\phi_k({\bf r}, t)$ and $n$ is the number of orbitals. For a molecule interacting with a time-dependent intense laser field the Kohn-Sham potential 
\begin{equation}
V_{KS}({\bf r}, t) = 
V_{ext}({\bf r},t) + 
\int \frac{\rho({\bf r}', t)}{|{\bf r}-{\bf r}'|} d{\bf r}' + 
V_{xc}({\bf r})
\end{equation}
includes the external potential due to the interaction of the electron with the $N$ nuclei in the molecule and with the time-dependent electric field: 
\begin{equation}
V_{ext}({\bf r},t) = 
\sum\limits_{i=1}^{N} -\frac{Z_i}{|{\bf R}_i-{\bf r}|} + 
E_0(t) \sin(\omega t) \sum\limits_{k=1}^{n} {\bf r}_k \cdot {\hat\epsilon}
\end{equation}
where $Z_i$ is the charge of the $i$th nucleus, $\hat\epsilon$ is the polarization direction, $\omega$ and $E_0(t)$ are the angular frequency and the time-dependent amplitude of the laser field. In the present calculations we considered a $\sin^2$-shaped envelope. 

The exact form of the exchange-correlation potential $V_{xc}$, which takes account of the multielectron effects, is unknown. To use TDDFT for practical calculations, different approaches have been proposed to design density functionals for the exchange-correlation energy (for an overview, see e.g., \cite{Gustavo2005}).  For the present calculations, we have performed systematic studies with various functionals and found that functionals based on the local density approximation (LDA), 
\begin{equation}
E_{xc}^{LDA}[\rho] = \int \rho({\bf r}) V_{xc}({\bf r}) d{\bf r} \; ,
\end{equation}
provide, in general, good results. An improvement is to take into account the correct asymptotic behavior ($1/r$), which can be done, for example, via the exchange-correlation potential proposed by van Leeuwen and Baerends \cite{Leeuwen1994}, 
\begin{eqnarray}
V_{xc}^{LB}(\alpha, \beta; {\bf r}) &=& 
\alpha V_x^{LDA}({\bf r}) + \beta V_c^{LDA}({\bf r}) 
\\
&& - \frac{\beta x^2({\bf r})\rho^{1/3}({\bf r})}
{1+3\beta x({\bf r})\ln[x^2({\bf r})+(x^2({\bf r})+1)^{1/2}]},
\nonumber
\end{eqnarray}
where $V_x^{LDA}$ and $V_c^{LDA}$ are the LDA exchange and correlation potentials and $x({\bf r}) = |\nabla \rho({\bf r})|/[\rho({\bf r})]^{4/3}$. $\alpha$ and $\beta$ are parameters obtained by fit to the exact exchange-correlation function of a certain atomic or molecular system. A similar TDDFT approach for the interaction of molecules with strong fields has been used recently by Chu and co-workers \cite{Chu2001a,Chu2001b}.

In order to solve the Kohn-Sham equations, Eq. (\ref{kohn}), we have discretized the wavefunction in space and time with uniform step $\Delta x = 0.03$ a.u. and $\Delta t = 0.03$ a.u., which converts the ansatz into a matrix equation using the Octopus code \cite{octopus2,octopus1}. The initial wavefunctions for the molecules considered in our study have been obtained by solving the eigenvalue problem self-consistently using an initial guess and geometry optimized using Octopus code as well (this ensures consistency and minimizes risk for errors). The wavefunction for each orbital is propagated forward in time using the enforced time-reversal symmetry method. We used grids that extend over 120 a.u. in polarization direction and 36 a.u. in the transverse directions. To suppress reflection of the wavefunctions at the boundary of the grid an imaginary absorbing potential has been applied.

\subsection{High-order harmonic generation from aligned molecules}

High-order harmonic generation is determined through the Fourier transform of the laser induced dipole moment in the target. Within the TDDFT formalism, the laser induced dipole moment is given by: 
\begin{equation}
{\bf d}_{tot} = \sum_{k=1}^n {\bf d}_k,
\end{equation}
where ${\bf d}_k$ is the contribution to the dipole moment from the $k$th Kohn-Sham orbital, 
\begin{equation}
{\bf d}_k =
\langle \phi_{k}({\bf r}, t)|{\bf r}|\phi_{k}({\bf r}, t) \rangle\, .
\end{equation}
The HHG spectrum is then found using the Fourier transform of the dipole moment, $d(\omega)$: 
\begin{equation}
P(\omega) = \frac{\omega^4}{12\pi\epsilon_0 c^3}{\bf d}(\omega) \cdot {\bf d}^*(\omega)\, .
\end{equation}
For the molecules studied below, the laser induced dipole moment has two components, parallel ($d_{||}$) and perpendicular ($d_{\perp}$) with respect to the direction of the electric field of the driving laser. The ellipticity of a given harmonic is then determined by: 
\begin{equation}
\epsilon = \sqrt{\frac{1+r^2-\sqrt{1+2r^2\cos(2\delta)+r^4}}{1+r^2+\sqrt{1+2r^2\cos(2\delta)+r^4}}}
\end{equation}
where 
\begin{equation}
r = \frac{|d_\perp(\omega)|}{|d_{||}(\omega)|}
\end{equation}
is the amplitude ratio and 
\begin{equation}
\delta = \arg[d_\perp(\omega)] - \arg[d_{||}(\omega)]
\end{equation}
is the relative phase between the two components. Maximum ellipticity, i.e.\ circular polarization, occurs for $r=1$ and $\delta = \pi$. 

\begin{figure}[t]
\centering\includegraphics[width=0.4\columnwidth]{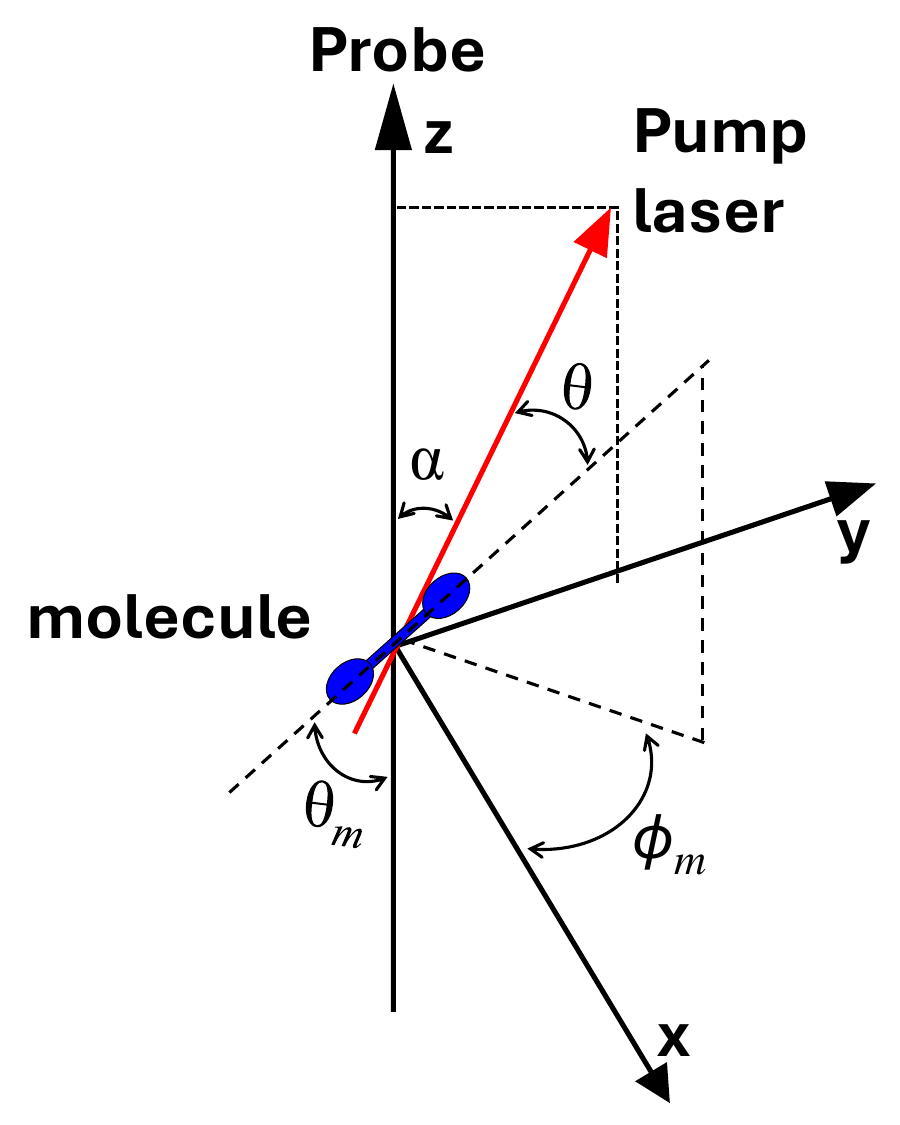}
\caption{\label{Fig:Average}Configuration of pump (aligning) pulse in the $y-z$ plane, probe (driver) pulse along the ${\hat z}$-direction and molecular axis.}
\end{figure}

In the experimental observations of the ellipticity in high-order harmonic generation of linear molecules, the molecules are often aligned by a pump laser pulse. The distribution of the alignment, achieved in the experiments, is typically measured via $\langle\cos^2(\theta)\rangle$, where $\theta$ is the angle between the polarization direction of the pump laser and the internuclear axis of the molecule (see Fig.\ \ref{Fig:Average}). In our simulations we have accounted for the experimental alignment of molecular ensemble by solving the Kohn-Sham equations for different alignment angles. For each angle, we obtained the parallel and perpendicular components of the dipole moment and then averaged them using the reported alignment distributions.

\section{Results}

In this section we present our results for the polarization and ellipticity of high-order harmonics from molecules H$_2^+$, H$_2$, and CO$_2$. The data for the different molecules provide us with the opportunity to compare our results with those from other theoretical analysis (for the one-electron system H$_2^+$) and demonstrate how multielectron effects and inner valence shell contributions influence the harmonics' ellipticity for the larger molecules. 

\subsection{Harmonic generation from H$_2^+$ and H$_2$}

\begin{figure}[t]
\centering
\includegraphics[width=0.4\columnwidth]{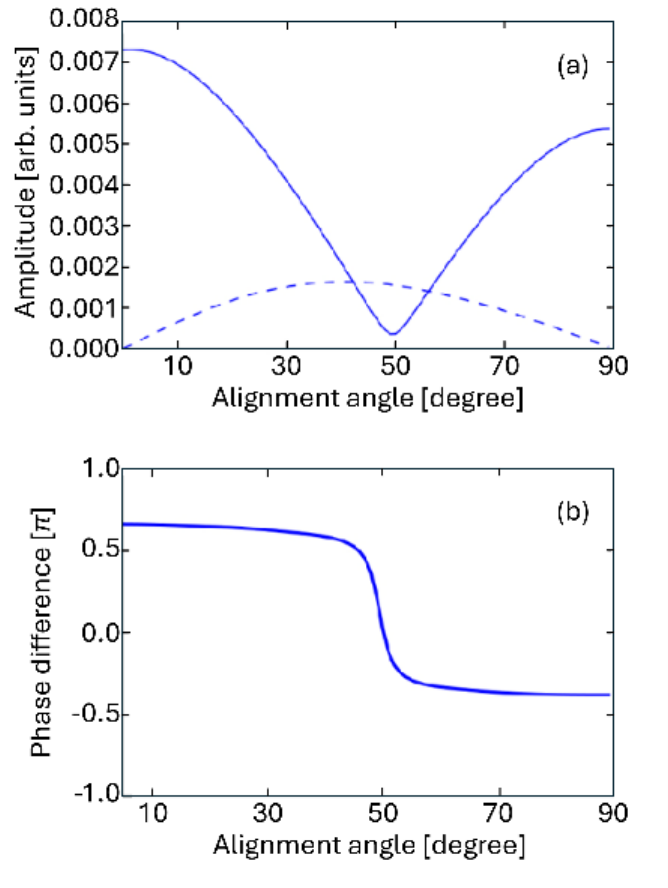}
\caption{Amplitudes of parallel and perpendicular components (a) and phase difference (b) of 57th harmonic order of H$_2^+$ as a function of the alignment angle. Laser parameters: 800 nm, $3 \times 10^{14}$ W/cm$^2$ and 30 fs.}
\label{Fig:H2plus}
\end{figure}

In order to test our numerical calculations, we first present results for the one-electron system H$_2^+$. In Fig.\ \ref{Fig:H2plus} we show results for the amplitudes (upper panel) and the phase difference (lower panel) for the 57th harmonics emitted from H$_2^+$ as a function of the alignment angle between the molecular axis and the polarization direction of a driving laser pulse at 800 nm and $3\times10^{14}$ W/cm$^2$ with a pulse duration of 30 fs. The laser parameters are chosen to be the same as in a recent work by Son et al. \cite{Son2010}, who studied the ellipticity of high-order harmonic generation from H$_2^+$ using the time-dependent generalized pseudospectral method. Our results are in good agreement with those previously reported for the overall shape of the components with a minimum at about $50^o$ for the parallel component and a phase jump at the same alignment angle. It has been shown before \cite{Le2010,vanderZwan2010,Son2010}, that these characteristic features are related to the two-center interference effect occurring in the parallel component.

\begin{figure}[t]
\centering\includegraphics[width=1\columnwidth]{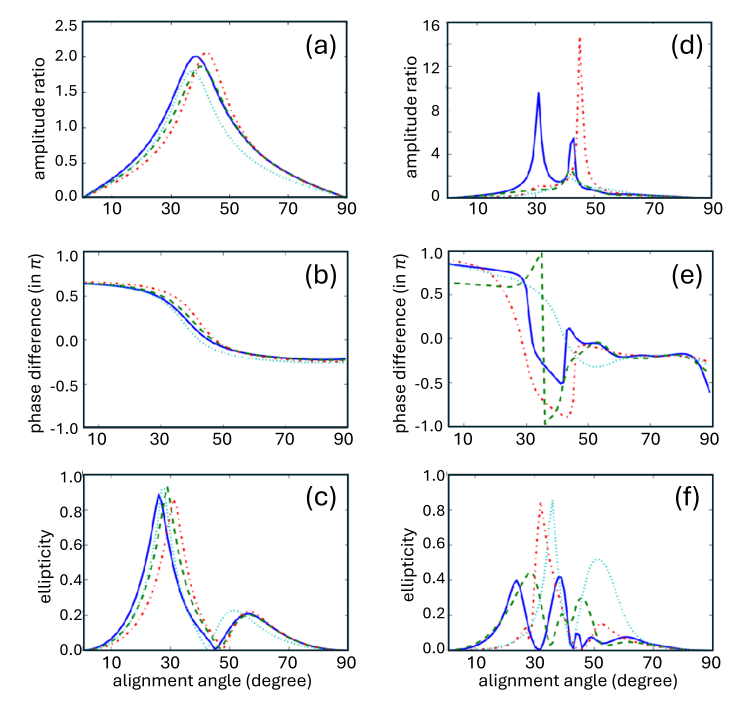}
\caption{Comparison of amplitude ratio $r$ (a, d), phase difference $\delta$ (b, e), and ellipticity (c, f) of high order harmonics from H$_2^+$ (a-c) and H$_2$ (d-f) as a function of the alignment angle: 27th (solid lines), 29th (dashed lines), 31st (dashed-dotted lines), and 33th harmonic (dotted lines). Laser parameters as in Fig. \ref{Fig:H2plus}.}
\label{Fig:H2plus-comp}
\end{figure}

In order to get an impression of the influence of multielectron effects on the ellipticity of high-order harmonics, we compare results for H$_2^+$ (Fig. \ref{Fig:H2plus-comp}, (a-c)) and H$_2$ (Fig. \ref{Fig:H2plus-comp}, (d-f)) obtained at the same set of laser parameters (800 nm, $3 \times 10^{14}$ W/cm$^2$). In each case we present theoretical predictions for four consecutive odd harmonics. For the single-electron molecule we observe, in agreement with our results in Fig.\ \ref{Fig:H2plus}, a maximum close to 1 in the ratio of the amplitude in parallel and perpendicular direction (a), a rapid change in the phase difference (b) and correspondingly a maximum in the ellipticity (c) around the alignment angle, at which the interference minimum in the specific harmonic occurs. For H$_2$, one would expect a similar pattern for the amplitude and the phase difference, since both electrons are in the same molecular orbital as in the case of H$_2^+$. Indeed, some features in the overall trend of the results in Fig.\ \ref{Fig:H2plus-comp} are similar, in particular we still note a maximum amplitude ratio (d) and a quick phase change (e) at about the same angles as for H$_2^+$. However, for the ratio we observe a much narrower structure and for the lowest harmonic a second maximum. On the other hand, the data for the phase difference are not as smooth as those for the single-electron molecule. As a result, we observe a much more complex pattern for the ellipticity of the harmonics generated from H$_2$ (f), although some maxima in the structures still occur near the alignment angle for the interference minimum. Thus, the comparison for the simplest molecules indicates that the ellipticity of high-order harmonics can be strongly influenced by multielectron effects. For larger molecules we may therefore expect even more complex features in the overall ellipticity patterns, since interferences from orbitals with different symmetry as well as coupling between different orbitals \cite{Xia2014,Xia2016} may play additional role.

\subsection{Harmonic generation from CO$_2$}

\begin{figure}[t]
\centering\includegraphics[width=0.5\columnwidth]{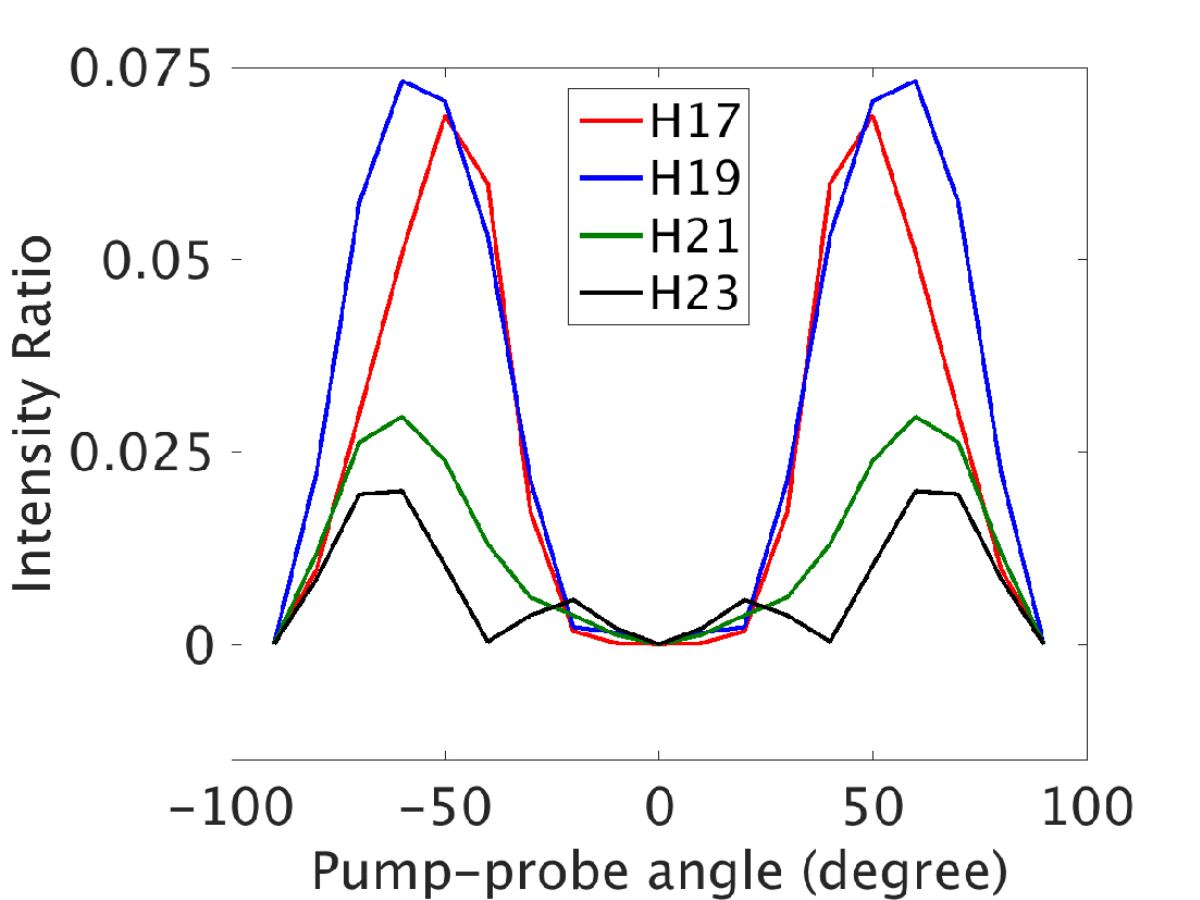}
\caption{
TDDFT results for the intensity ratio of perpendicular to parallel component of four consecutive harmonics in CO$_2$ as a function of the angle between the pump and the 30 fs probe laser pulse at 800 nm and $1.5 \times 10^{14}$ W/cm$^2$: 17th (red line), 19th (blue line), 21st (green line) and 23rd harmonic (black line). For each angle, the experimental reported alignment distribution \cite{Zhao2009} was considered in the calculations.}
\label{Fig:CO2-comparison}
\end{figure}

Next, we analyze the results of our calculations for the ellipticity in the harmonic generation from the more complex but linear triatomic molecule CO$_2$, which has been also studied experimentally \cite{Zhao2009}. In order to compare with the experimental data, we have obtained the intensity ratio of the perpendicular to parallel component of the harmonic emission as a function of the angle between the pump and probe laser pulse. For each orientation angle considered, we have taken into account the experimentally reported alignment distribution by performing an average over the simulation results for the respective alignment angles in the distribution. Our results in Fig.\ \ref{Fig:CO2-comparison} show a rather small intensity ratio and, hence, relatively small ellipticity  with a maximum at about a relative angle of about $60^o$ between polarization direction of pump and probe pulse for each of the harmonics studied experimentally. The absolute values as well as the position of the maxima are in very good agreement with the observations by Zhao et al. \cite{Zhao2009}. The observed and calculated rather weak perpendicular component of the harmonics in CO$_2$ is in contrast to results for N$_2$, for which both experiment \cite{Zhao2009,Mairesse2010} and TDDFT \cite{Xia2014} as well as other calculations \cite{Mairesse2010,Le2010,Son2010} show a strong ellipticity for the emitted harmonics at certain alignment angles. 

\begin{figure}[t]
\centering
\includegraphics[width=0.49\columnwidth]{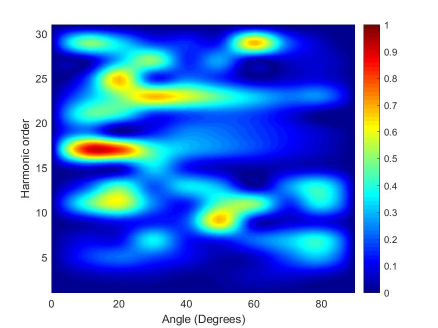}
\includegraphics[width=0.49\columnwidth]{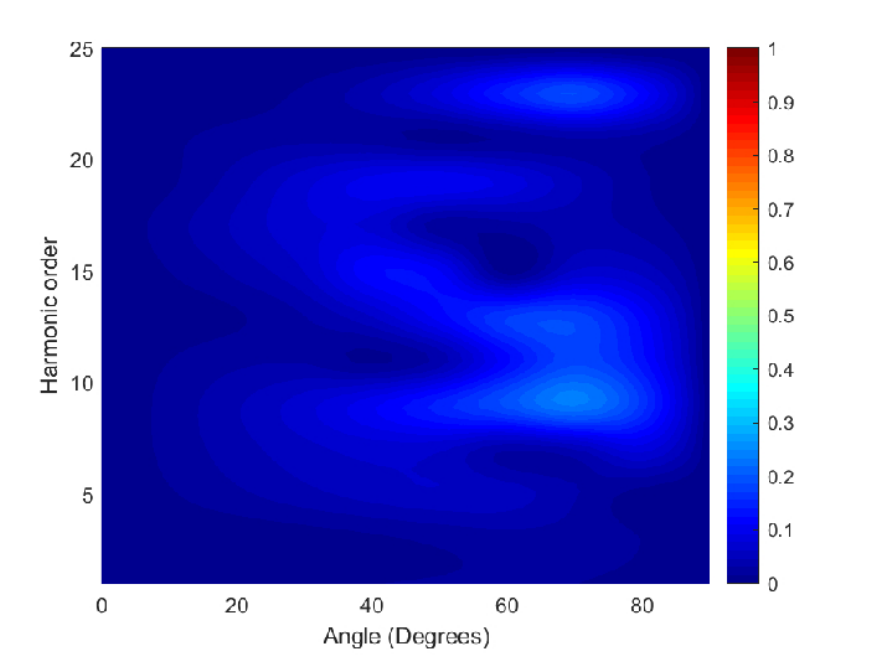}
\caption{
Comparison of results for the ellipticity of high-order harmonics as a function of alignment angle for CO$_2$: without (a) and with averaging (b).
Laser parameters as in Fig.\ \ref{Fig:CO2-comparison}.}
\label{Fig:CO2-averaging}
\end{figure}

Part of the explanation for the weak ellipticity is due to the ensemble angle average effect, which reduces the overall ellipticity, as observed before in N$_2$ \cite{Xia2014}. The effect can be seen from the comparison of the harmonics ellipticity as a function of the alignment angle without (a) and with (b) average in Fig. \ref{Fig:CO2-averaging}. It is clearly seen that, in particular for the lower-order harmonics (below 15th harmonics), without averaging there is a strong ellipticity for certain alignment angles which disappears after alignment average is taken into account. In contrast, for the experimentally reported data in the range of 17th to 23rd harmonics the averaging process does have a smaller effect only.

\begin{figure}[t]
\centering\includegraphics[width=1\columnwidth]{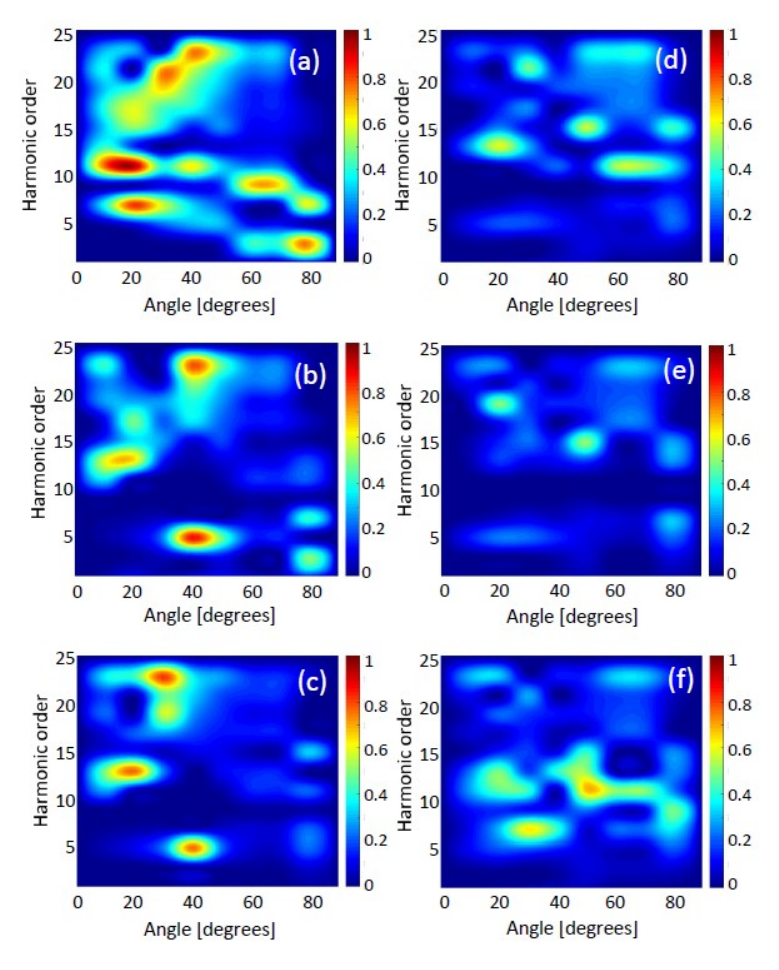}
\caption{Ellipticity of high-order harmonics as a function of alignment angle for CO$_2$. Starting with the results from HOMO only (a): $(1\pi_g)^4$, contributions from inner valence orbitals are added subsequently in the other panels: (b) $(3\sigma_u)^2(1\pi_g)^4$, (c) $(1\pi_u)^4(3\sigma_u)^2(1\pi_g)^4$, (d) 
$(4\sigma_g)^2(1\pi_u)^4(3\sigma_u)^2(1\pi_g)^4$, (e) 
$(2\sigma_u)^2(4\sigma_g)^2(1\pi_u)^4(3\sigma_u)^2(1\pi_g)^4$, (f) $(3\sigma_g)^2(2\sigma_u)^2(4\sigma_g)^2$
$(1\pi_u)^4(3\sigma_u)^2(1\pi_g)^4$.  Laser parameters as in Fig.\ \ref{Fig:CO2-comparison}.}
\label{co2noav}
\end{figure}

\begin{figure}[t]
\centering\includegraphics[width=1\columnwidth]{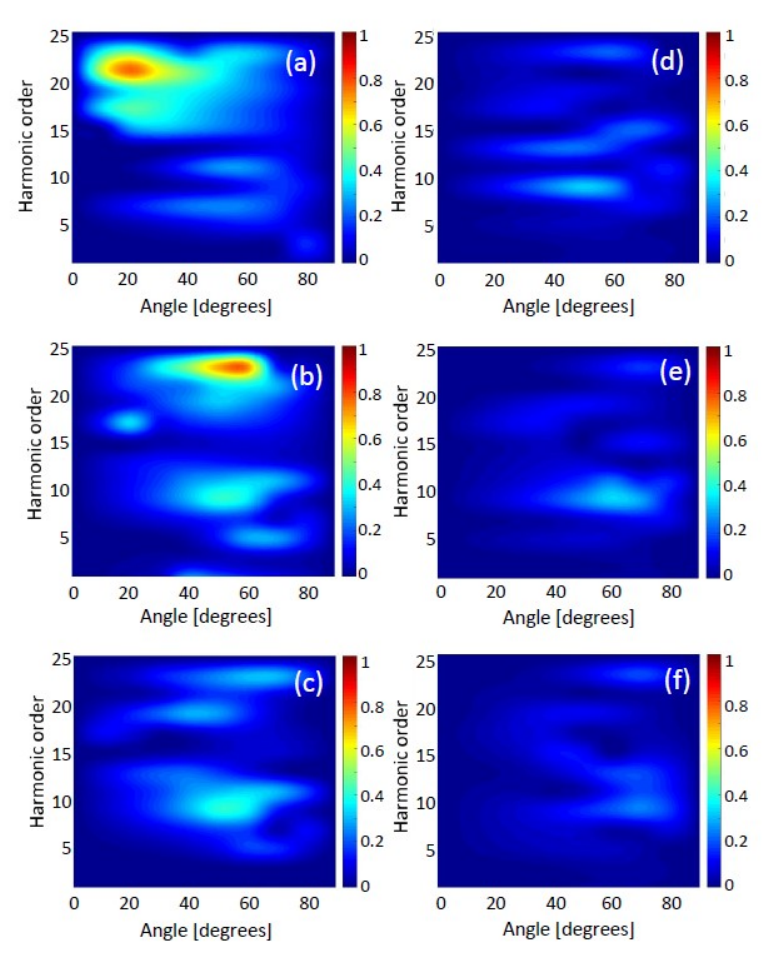}
\caption{Results after rotational averaging.
The rest of the notation and parameters as in fig.\ref{co2noav}. }
\label{co2cos4}
\end{figure}

In this latter range of harmonics from CO$_2$ the main origin for the weak ellipticity is actually the role of multielectron effects involving contributions from several orbitals. In order to analyze these contributions, we compare in Fig.\ \ref{co2noav} the ellipticity of the harmonic response from the HOMO only (a) with those when adding subsequently the contributions from the inner valence orbitals up to HOMO-5 (f). The comparison shows that the ellipticity of high-order harmonics from CO$_2$ is influenced by the six valence orbitals considered. While the ellipticity of 17th to 23rd harmonics generated from the HOMO is rather large for certain alignments angles, the ellipticity gradually gets weaker as more contributions are added. In contrast, for harmonics around the cutoff there remains a strong ellipticity at some alignment angles.

The ellipticity of higher-order harmonics at certain alignment angles from the HOMO ($3 \pi_g$) can be understood based on the two-center interference effect, similar as in the case of H$_2^+$ and H$_2$ above. The importance of such orbital structure effect for the harmonic generation from the HOMO of CO$_2$ has been pointed out before \cite{Le2010}. The strong contributions from the inner valence orbitals originate on a variety of effects. Both, HOMO-1 ($2 \sigma_u$) and HOMO-2 ($1 \pi_u$) have a different orbital symmetry than the HOMO of CO$_2$. Therefore, ionization and, hence, harmonic generation, from HOMO is suppressed due to destructive interference at alignment angles of $0^\circ$ and $90^\circ$ while it is at maximum around $45^\circ$ \cite{Le2009PRA}. In contrast, the ionization rate is largest at $0^\circ$ for HOMO-1 and $90^\circ$ for HOMO-2. Consequently, high-order harmonic generation from these two orbitals contributes strongly close to alignment angles at which the signal from the HOMO is weakest, despite the fact that the ionization potential for the inner valence orbitals is smaller than that of the HOMO. 

When we include in the calculations the experimentally reported alignment distribution and performed rotational averaging over the simulation results for the respective alignment angles in the distribution one can see in Fig.  \ref{co2cos4} that the overall changes in the ellipticities of harmonics when adding subsequent contributions from orbitals are rather small. Each subsequent partial orbital contribution leads to effects of destructive interference, which after comparing to Fig. \ref{co2noav} seems to be even more enhanced by rotational averaging. After adding subsequent orbitals to the contributions of the first 3, as can be seen in Figs. \ref{co2cos4} (d)-(f), the rotationally averaged HHG ellipticities are very small and remain small, while only very little changes can be seen when considering modifications due to contributions from deeper lying orbitals. 

\begin{figure}[t]
\centering\includegraphics[width=0.4\columnwidth]{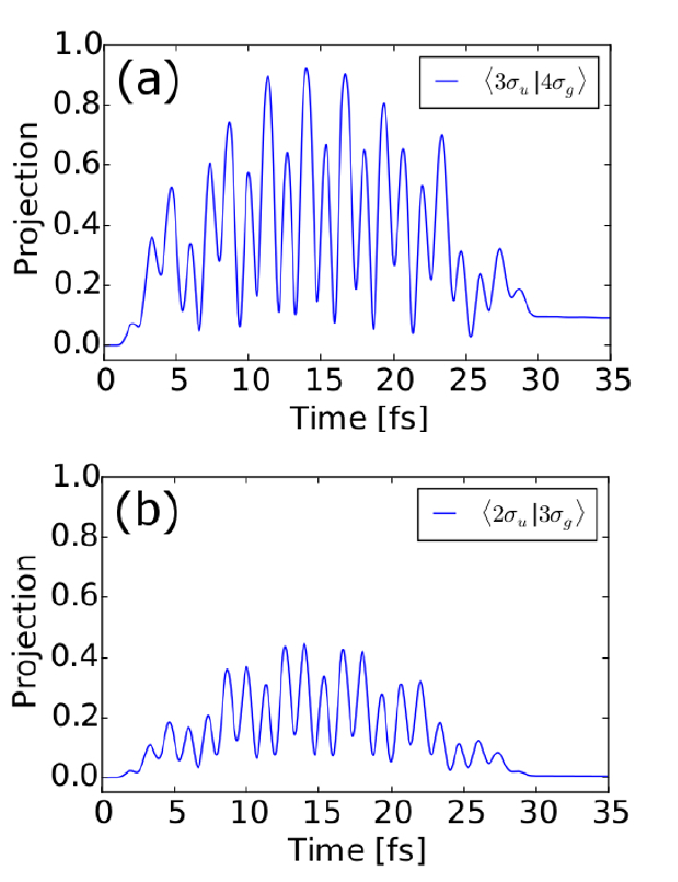}
\caption{
Projection of coupled inner valence orbitals (a) HOMO-3 ($4 \sigma_g$) to HOMO-1 ($3 \pi_u$) (a) and (b) HOMO-5 ($3 \sigma_g$) to HOMO-4 ($2 \pi_u$) (b) for an alignment angle of $20^o$. Laser parameters as in Fig.\ \ref{Fig:CO2-comparison}.
}
\label{Fig:CO2-projections}
\end{figure}

Regarding  the contributions form the deeper lying inner valence orbitals, we have found that these are either strongly coupled to one of the higher lying states or among each other by the driving field. In the case of the HOMO-3 state ($2 \sigma_g$), the projection onto the HOMO-1 state is shown in Fig.\ \ref{Fig:CO2-projections}(a). We observe a strong coupling driven by the field although the frequency is non-resonant. This explains the significant change in the ellipticity pattern upon inclusion of the HOMO-3 state (Fig.\ \ref{co2noav}(d)). Finally, HOMO-4 and HOMO-5 states slightly contribute to the 17th to 23rd harmonic generation at the given parameters and, hence, to the ellipticity pattern, since these two orbitals are coupled with each other, leading to a population transfer of about 40\% (see Fig.\ \ref{Fig:CO2-projections}(b)). 

\subsection{Summary}

To summarize, our results obtained within the time-dependent density functional theory indicate that high-order harmonic generation from CO$_2$ is influenced by multielectron effects with contributions from a significant number of inner-valence orbitals, besides the contribution from the HOMO. The harmonic emission from these orbitals is strongest at different alignment angles due to interference effects arising from the specific orbital structures and there is a strong laser driven coupling between certain orbitals. As a result,  the overall ellipticity of the higher-order harmonics is rather small, except for the cutoff harmonics. The partial alignment and the related averaging of the results for different orientation angles further diminishes the ellipticity.  

\section*{Acknowledgments}

A.J. acknowledges support from  
 the U.S. National Science Foundation (Grant No.  PHY-231714 and Grant No.
PHY-2110628). This work utilized the Alpine high-performance computing resource at the
University of Colorado Boulder. Alpine is jointly funded by the University of Colorado Boulder, the University of Colorado Anschutz, and Colorado State University.

\end{document}